# Physical learning in reprogrammable metamaterials for adaptation to unknown environments

Kai Jun Chen[1], Catherine Catrambone[1], Christopher Sowinski[1], Jacob Mukobi[1], Enzo Andreacchio[2], Enquan Chew[2], Alexandre Morland[1], Maria Sakovsky[1]

[1]Department of Aeronautics and Astronautics, Stanford University, 496 Lomita Mall, Stanford 94305, CA, USA

[2]Department of Mechanical Engineering, Stanford University, 440 Escondido Mall, Stanford 94305, CA, USA

**Abstract**

Reprogrammable mechanical metamaterials, composed of a lattice of discretely adaptive elements, are emerging as a promising platform for mechanical intelligence. To operate in unknown environments, such structures must go beyond passive responsiveness and embody traits of mechanical intelligence: sensing, computing, adaptation, and memory. However, current approaches fall short, as computation of the required adaptation in response to changes in environmental stimuli must be pre-computed ahead of operation. Here we present a physical learning approach that harnesses the structure's mechanics to perform computation and drive adaptation. The desired global deformation response of nonlinear metamaterials with adaptive stiffness is physically encoded as local strain targets across internal adaptive elements. The structure adapts by iteratively interacting with the environment and updating its stiffness distribution using a model-free algorithm. The resulting system demonstrates autonomous real-time adaptation (~seconds) to previously unknown loading conditions without pre-computation. Physical learning inherently accounts for manufacturing imperfections and is robust to sensor noise and structural damage. We also demonstrate scalability to complex metamaterial structures and different metamaterial architectures. By uniting sensing, computation, and actuation in a mechanical framework, this work makes key strides towards embodying the traits of mechanical intelligence into adaptive structures. We expect our approach to open pathways towards in-situ adaptation to unknown environment for applications in hypersonic flight, adaptive robotics, and exploration in extreme environments.

**Main article**

Responsive mechanical metamaterials react to thermal[1–3], light[4,5], mechanical[6–8], and electromagnetic[9–11] environmental stimuli in prescribed ways that are fixed through the design and fabrication. They have enabled pre-programmed untethered locomotion control[12], stiffness tuning[13], multi-stable shape change[14], and antenna reconfiguration[15]. To enable operation in dynamic and unknown environments, metamaterials must embody mechanical intelligence to sense the environment, compute the required mechanical changes, adapt mechanical properties, and remember learnt behaviors[16]. For example, Xia et al.[17] hypothesized that mechanical metamaterials could exhibit neuromorphism – the ability to learn from past and present environmental interactions and adapt their mechanical properties online by emulating the structure of biological neural networks. The hypothesis is rooted in the analogy between heterogeneously- and time-varying mechanical properties in an architected geometry and synaptic weights in the brain.

To support this vision, recent efforts have demonstrated reprogrammable adaptation post-fabrication using lattices of elements, each with adaptive mechanics, to modulate stiffness, deformation pathways, or load-bearing capability[18–21]. This functionality relies on a large number of degrees of freedom for adaptation and hence a large state space. For example, Mechanical Neural Networks[22,23] are a mechanical analogue to biological neural networks, with stiffness adaptation of elements analogous to synaptic weights[24].

Current adaptive mechanical metamaterials fall short of the hallmarks of mechanical intelligence and neuromorphism. The main challenge lies in computing the required mechanical adaptation given a sensed environmental stimulus. This presents a non-convex, ill-posed, and discrete inverse problem typically solved via established gradient-free optimization algorithms[20,22]. As such, the adaptation is pre-computed ahead of operation, preventing a real-time sense-assess-respond loop.

Unconventional computing approaches inspired by principles of information processing in physical systems[25,26], such as physical reservoir computing[27–30], mechanical computing[31–34], and bio-inspired computing[24,35–37], are a promising means of transforming the metamaterial itself into a computational resource. Mechanics offers practically instantaneous computation[38], operation in extreme environments[32], and cybersecure systems[38]. However, leveraging embodied physics-based interactions of the structure with its environment for computing the required adaptation actions remains unexplored.

We propose a physical learning approach for driving adaptation in reprogrammable mechanical metamaterials (Fig. 1a). The mechanics of the structure process environmental stimuli into measurable physical quantities that inform adaptation. The adaptive structure iteratively interacts with an unknown environment, extracts its own state via distributed sensors, and adapts its mechanical properties following a model-free algorithm until convergence to a desired mechanical response. We demonstrate that adaptive structures learn target shape adaptation and force control strategies in a handful of iterations, enabling response to unknown environments in real-time (~seconds). We present the first reprogrammable mechanical metamaterial to autonomously adapt mechanical performance to unknown loading in real-time without pre-computing the adaptation (Fig. 1b). Our approach does not require solution of an optimization problem or data-driven training, is deterministic with predictable accuracy and learning rates, and is computationally simple, relying on very few algebraic operations. Leveraging mechanics for computing also inherently accounts for manufacturing imperfections without the arduous task of characterizing them. Our work makes key strides towards embodying the traits of mechanical intelligence into adaptive metamaterials. Real-time adaptation is an enabling feature in adaptive systems for highly dynamic environments including turbulent and hypersonic aircraft flight, vibration isolation, and robotic gripping of unknown objects (Fig. 1a).

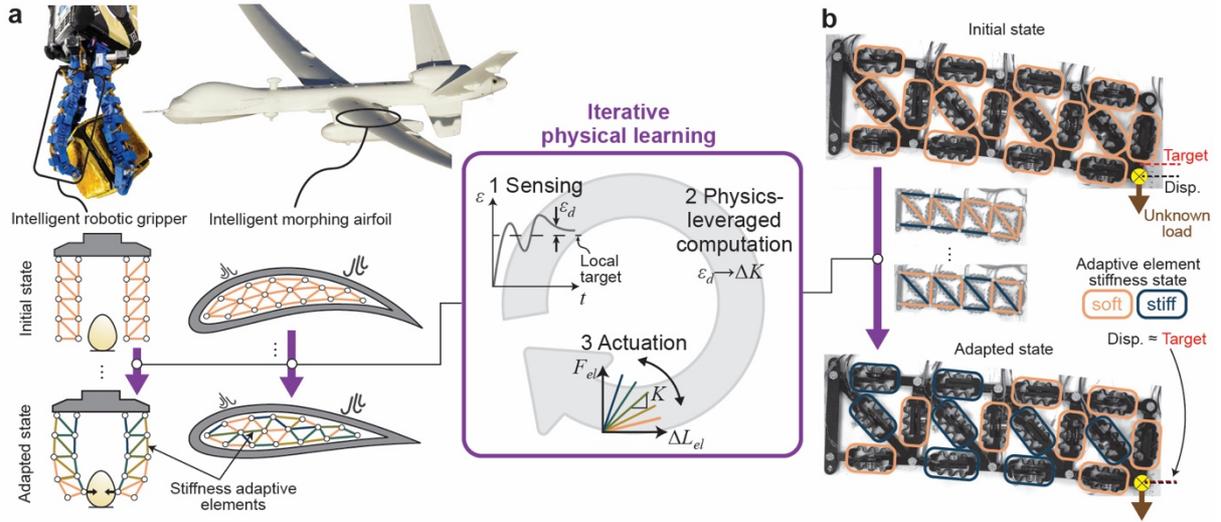

**Fig. 1 | Overview of Physical Learning Approach. a.** (Right) Proposed iterative physical learning approach for a reprogrammable mechanical metamaterial consisting of adaptive elements with sensing and actuatable stiffness states, driven by a model-free algorithm for physics-leveraged computation. (Left) Envisioning potential applications of the physical learning approach to enable mechanically intelligent adaptive systems. Examples include robotic grippers for optimizing grasp of unknown objects and morphing airfoils for wind gust rejection. Image sources: space robotic gripper (NASA, 2025); unmanned aerial vehicle (NASA, 2006). **b.** Experimental metamaterial beam structure, under an unknown external tip load, autonomously adapts toward a target tip displacement by applying the physical learning approach through learning encounters with the load. Stiffness configurations for the metamaterial beam in the initial ($k=0$) and adapted ($k_s=6$) states in experimental testing for unknown load $F_{ext}$=15 N, target tip displacement $d_t$=28 mm.

## Iterative physical learning

We consider a class of mechanical metamaterials composed of adaptive elements, each capable of modifying stiffness between some discrete states. To demonstrate our physical learning approach, we use an extension-dominated lattice in the form of a cantilever beam (Fig. 1b). Each bar has discretely reprogrammable stiffness and integrated strain sensors. We consider a high deformation problem with geometric non-linearities commonly found across a variety of engineering disciplines. More complex structures are also studied in subsequent sections. Here, the metamaterial beam is tasked with maintaining a target tip displacement, $d_t$, under some unknown tip load, $F_{ext}$. To perform this adaptation, the beam must modify the stiffness configuration, $\mathbf{K}$, of its adaptive elements, i.e.,

$$\text{Find } \mathbf{K} = \{K_1, K_2, \ldots, K_n\} \text{ where } K_i \in \{K^{(1)}, \ldots, K^{(m)}\} \text{ s.t. } |(d(\mathbf{K}) - d_t)/d_t| < \delta, \qquad (1)$$

where $n$ is the number of adaptive elements in the lattice, $K_i$ is the stiffness of the i$^{\text{th}}$ adaptive element, which has $m$ available discrete states, $d$ is the actual tip displacement, $d_t$ is the target tip displacement, and $\delta$ is a user-prescribed error tolerance denoting the target band.

To find a suitable stiffness solution, $\mathbf{K}$, we introduce a model-free decision-making algorithm that locally minimizes strain deviations from a target shape as the structure interacts iteratively with its environment. In our biology analogy, this is the equivalent to adjusting synaptic weights in the brain as we learn new tasks. First, we establish target strains for each adaptive element, $\varepsilon_{t,i}$, by applying the target tip displacement, $d_t$, in some arbitrary stiffness configuration (e.g., all adaptive elements in their lowest stiffness state, $K^{(1)}$). This associates the internal states of the metamaterial with the desired global response, using the structure itself to embody the required information for adaptation. This initialization only needs to be performed once and can be stored in the structure with a small memory requirement. During operation, the beam is subjected to an unknown tip load, $F_{ext}$, and the resulting element strains, $\varepsilon_{c,i}$, are measured. Each element stiffness is updated to locally minimize strain deviations, $\varepsilon_{d,i}=|\varepsilon_{c,i}-\varepsilon_{t,i}|$. If the element's current strain magnitude exceeds its target, i.e. $|\varepsilon_{c,i}|>|\varepsilon_{t,i}|$, then it increases its stiffness to the next discrete state, except if it is already at the maximum stiffness. Conversely, the element is softened if $|\varepsilon_{c,i}|<|\varepsilon_{t,i}|$ and it is not already in the minimum stiffness state. Stiffness is updated iteratively until the tip displacement, $d(\mathbf{K})$, is within $\pm\delta$ of the target tip displacement, $d_t$.

We limit the number of actuations per iteration to $n_c$, prioritizing elements with the largest strain deviations, $\varepsilon_{d,i}$. Simultaneous actuation of all elements at each iteration leads to an unstable response. We find that progressively adjusting $n_c$ greatly improves learning rate, by allowing more actuations when the error is initially large, then subsequently reducing the number of actuations for precise control. Algorithmically, we regulate $n_c$ according to:

$$n_c = \min\{\text{round}(np_0\gamma^k), 1\}, \tag{2}$$

where $p_0 \in (0,1]$ is the proportion of adaptive elements that actuate in the first iteration, $\gamma \in (0,1]$ is a decay parameter, and k is the iteration. Physical learning is terminated only once the tip displacement error $|(d-d_t)/d_t|$ is below $\delta$. We define a successful adaptation as one where the tip displacement reaches the target band within 100 iterations (arbitrarily chosen). Finally, to assist the metamaterial to overcome local minima in the rugged nonconvex solution space, we introduce short-term memory that permits an element to actuate only if it did not actuate in the previous iteration (see Methods section 'Model-free algorithm' for supporting data). A detailed flowchart of the algorithm is provided in Extended Data Fig. 1. Overall, our physical learning approach uses deformation in nonlinear metamaterials to iteratively process environmental stimuli and inform actuation decisions, thereby leveraging the structure's mechanics as a form of physical computing.

**Characterizing learning rates and accuracy**

To demonstrate physical learning, we first consider a simple metamaterial beam with $n=16$ adaptive elements arranged in an extension-dominated lattice (Fig. 2a), each with $m=2$ stiffness states ($K_i \in \{13, 91\}$ N/mm). The mechanical response of the structure was simulated in Python using a finite element code that models the metamaterial beam as a geometrically non-linear collection of linear elastic bars (see Methods section 'Simulation of metamaterial beam'). For the adaptation task, we prescribe $d_t = 28$ mm, to provide a geometrically non-linear problem, and a target band of $\delta = 10\%$. We assign an initial, untrained configuration of all elements in the soft state (i.e., all $K_i = K^{(1)}$). Hyper-parameters $p_0 = 0.31$, $\gamma = 0.7$ (Eq. 2) are used.

Physical learning is tested for a range of nine $F_{ext}$ values from 11 N to 43 N. This represents the range of external forces for which the lattice can physically adapt to the prescribed $d_t$. For all $F_{ext}$ values, we find that the tip displacement is initially far from target due to the unanticipated external load and adapts to the target band within 9 iterations, at which point the learning process is terminated (Fig. 2a). Furthermore, in each case, the metamaterial beam adapts using the same target strains, and there is no need to reinitialize new targets for a different load.

We quantify the learning performance of the beam in Fig. 2a using two metrics: learning rate, $k_s$, defined as the first iteration when $|(d-d_t)/d_t|<\delta$, and the accuracy, $e_s$, defined as the error $e=|d-d_t|/d_t$ when $k=k_s$. Averaging across these nine external loading cases, we obtain $\bar{k}_s$=4.1 and $\bar{e}_s$=2.2%. Additionally, the worst-case performance for this structure is $k_{s,max}$=9 and $e_{s,max}$=9.3% and occurs for a load of ….

The choice of $p_0$ and $\gamma$ is key to obtain successful adaptation. For $\delta$=10%, we show which hyper-parameters result in successful adaption across all nine external loads (green in Fig. 2b). The parameters used in Fig. 2a are indicated with a star. We observe a general clustering of successful parameter combinations in two regions, one roughly in the bottom-left of a diagonal line and another roughly centered about the top-right. This offers a heuristic for selecting appropriate hyper-parameters to enable the algorithm to successfully traverse the highly non-convex, rugged solution space.

Varying $p_0$ and $\gamma$ also affects the learning rate and accuracy. A systematic exploration of the parameter space, varying $\delta$ from 5 to 20%, and considering only successful hyper-parameters, allows us to generate a Pareto front (Fig. 2c) denoting optimal hyper-parameters that trade-off learning rate and accuracy. The horizontal and vertical asymptotes of the Pareto-front represent the maximum achievable learning rate and accuracy respectively. Notably, the values $p_0$ = 0.31, $\gamma$ = 0.7 used above are Pareto-optimal (Fig. 2c), leading to an adaptation that resembles a 'critically-damped' response (Fig. 2a). On the other hand, allowing one actuation per iteration (i.e., $n_c$=1 for all $k$ using $p_0$ = 0.06, $\gamma$ = 1), is not Pareto-optimal. This still yields successful adaptation but with slower learning ($\bar{k}_s$ = 7.7) and reduced accuracy ($\bar{e}_s$ = 4.4%). We see that appropriate selection of hyper-parameters allows physical learning to adapt in <10 iterations, with an accuracy as good as 2% for this particular structure.

We find that increasing the number of degrees-of-freedom (DOFs) can improve achievable learning accuracy further. For adaptive elements with two stiffness states, $m$=2, increasing the number of adaptive elements (see Methods section 'Parametric study of lattice parameters') shifts the Pareto front to the left (Fig. 2d), with accuracy of $\bar{e}_s$ up to ~0.4% possible. Furthermore, the rapid learning rate is preserved even for very large $n$. On the other hand, for a fixed $n$=16, increasing the number of available stiffness states shift the Pareto front leftward and upward (Fig. 2e). Here, the accuracy continues to improve with an increasing number of stiffness states, without the previous limit. However, we observe slower learning rates, as the algorithm must step through more stiffness states for each element. Overall, these observations provide practical design guidelines on constructing more complex metamaterials. Structures with more DOFs improve learning accuracy. However, the experimental implementation can be complex due to the need for miniaturizing elements and increasing number of stiffness states. Nevertheless, recent trends in mechanical microfabrication of stiffness-adaptive elements[32] and the design of highly multi-stable elements[39,40] could make high DOF metamaterials realistic.

Physical learning demonstrates a faster learning rate ($k_s$ = 3~8) (for $n$=16, $m$=2) than model-free optimization algorithms such as partial pattern search (PPS) ($k_s$=24~74) and genetic algorithm (GA) ($k_s$=51~420) (see Supplementary 1 'Benchmark with Conventional Optimization Algorithms'). This highlights the ability of physical learning to operate online, while conventional algorithms would require pre-computing a solution. Additionally, the deterministic nature of physical learning allows us to place guarantees on adaptation performance, in contrast to the inherent stochasticity of PPS and GA. However, for lattices with fewer DOFs ($n$ < ~208 or $m$ < ~5), physical learning cannot access exceedingly accurate solutions ($\bar{e}_s$<1%). We know that such accurate solutions are possible as found using GA, and so this accuracy limitation arises from the search dynamics of the algorithm trapping it in local minima.

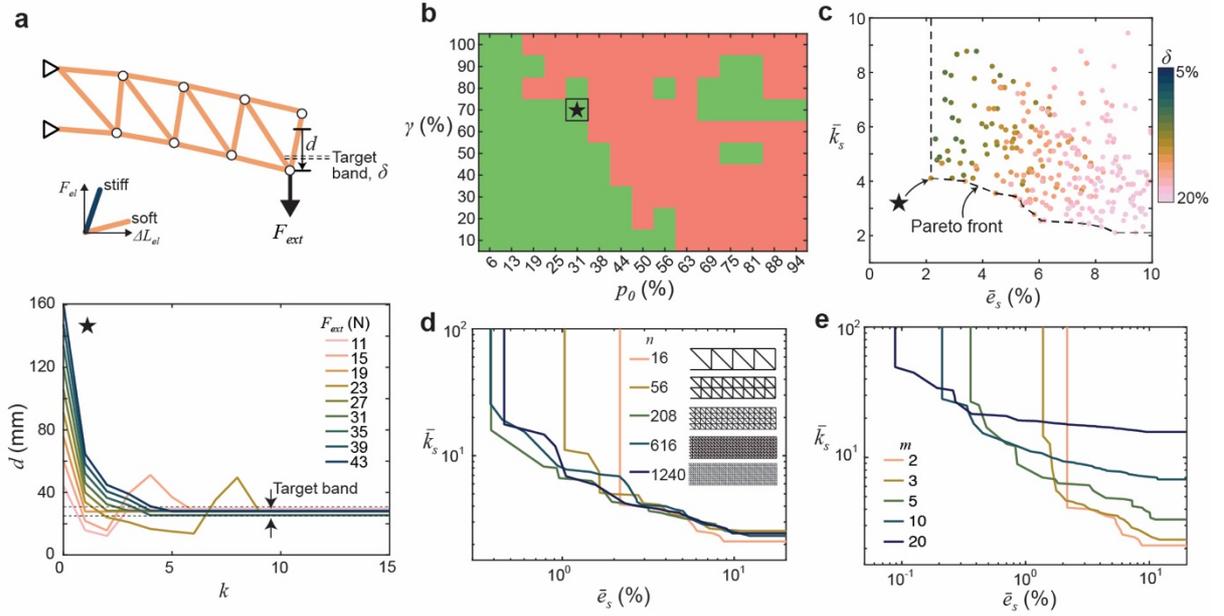

**Fig. 2 | Tuning learning rate and accuracy in extension-dominated adaptive metamaterial beam. a.** Iterative adaptation of the tip displacement via physical learning across various load test cases for $n$ = 16, $m$ = 2. **b.** Hyper-parameter combinations that successfully adapt within $\delta$ =10% across all nine $F_{ext}$ values (from Fig. 2a legend) indicated in green and failed ones in red. The cell with a star denotes the hyper-parameters used in Fig. 2a. **c.** Learning rates, $\bar{k}_s$, and accuracy, $\bar{e}_s$, as a function of hyper-parameters, $p_0$ and $\gamma$, and target bands, 5% < $\delta$ < 20%. Each point denotes a successful hyper-parameter combination. The point labelled with a star corresponds to the result from Fig. 2a. The dotted line indicates the approximate Pareto front, representing a trade-off between learning rate and accuracy. **d.** Effect of increasing the number of adaptive elements, $n$, on the Pareto front for a fixed number of stiffness states, $m$=2. **e.** Effect of increasing the number of available stiffness states per element, $m$, on the Pareto front a fixed number of adaptive elements, $n$=16.

**Experimental demonstration of autonomous adaptation via physical learning**

We provide the first demonstration of autonomous real-time learning in an adaptive structure with integrated sense-assess-respond functionality (Fig. 1b). We apply physical learning to a metamaterial beam demonstrator composed of $n$=16 adaptive elements, each with $m$=2 stiffness states (Fig. 3a). In the soft state ($K^{(1)}$=13 N/mm), the load is carried through compliant circular arches, whereas in the stiff state ($K^{(2)}$=91 N/mm), it is carried via an interlocking central column (Fig. 3b). Magnetic actuation is used to switch stiffness states. The actuation is powered for only ~100 ms and each stiffness state can be maintained passively through reversible latches

(see Supplementary 2 'Video of adaptive element'). Strain gauges are mounted to the compliant arches of every element to provide in situ sensing. Finally, the structure is connected to an Arduino controller board to manage the actuations (see Methods section 'Reprogrammable metamaterial testbed' for details on design, fabrication and operation). Notably, due to the simplicity of the model-free algorithm, the experimental implementation requires only simple electronics, with the decision-making process arising primarily through the physical interaction of the metamaterial with the applied load.

Across three test cases of $F_{ext} \in \{11,15,19\}$(N) (with $d_t$=28 mm, $\delta$=10%), we see that the tip displacement consistently adapts to the target band in experiments (Fig. 3c, see Supplementary 3 'Video of physical learning' for $F_{ext}$=15 N case). Using the Pareto-optimal hyperparameters $p_0$ = 0.31, $\gamma$ = 0.7, we obtain the performance metrics of $\bar{k}_s$=3.3 and $\bar{e}_s$=4.9%, demonstrating rapid adaptation capability with a reasonably small error considering experimental errors. Our simulations predict $\bar{k}_s$=3.3 and $\bar{e}_s$=2.1%. In addition, the tip displacement response in experiments closely matches the simulation (Fig. 3c). Each iteration of sensing and actuation takes just 2~6 s in our demonstrator, with complete adaptation achieved in 8~25 s excluding loading time.

Interestingly, we find that the sequence of actuations differs in experiment and simulation. Fig. 3d shows the stiffness distribution at every iteration for the experiment and simulation for $F_{ext}$ = 15 N. The order of actuations in each iteration is denoted by the roman numerals. For the first iteration, $k$ = 1, the same elements were actuated but in a different order, while for the second iteration, $k$ = 2, different elements were actuated. The final stiffness distribution in experiments and simulations is also different, while the overall tip displacement target is met in both. Manufacturing imperfections creating stiffness variations across adaptive elements, friction in the lattice assembly, and sensor noise are responsible for differences in the in situ measured strain deviations, $\varepsilon_{d,i}$, compared to the simulation. Physical learning therefore selects different elements to actuate. These actuation decisions are made with imperfections accounted for without the need to explicitly characterize them. The only factor not captured in the structure's mechanics is sensor noise. We characterize this experimentally and find that it has no measurable effect on the adaptation response. Furthermore, artificially increasing sensor noise by up to 3 orders of magnitude in simulation still results in successful adaptation with some decrease in learning rate (see Supplementary 4 'Characterization of sensor noise' for details). As such, sensor noise has little influence in our experiments, and we attribute deviations to manufacturing imperfections. The fact that imperfections are inherently captured in the mechanics of the adaptive structure is a key advantage of physical learning. Selection of Pareto-optimal hyper-parameters allows users to achieve predicted learning rates and accuracy experimentally, even with manufacturing imperfections.

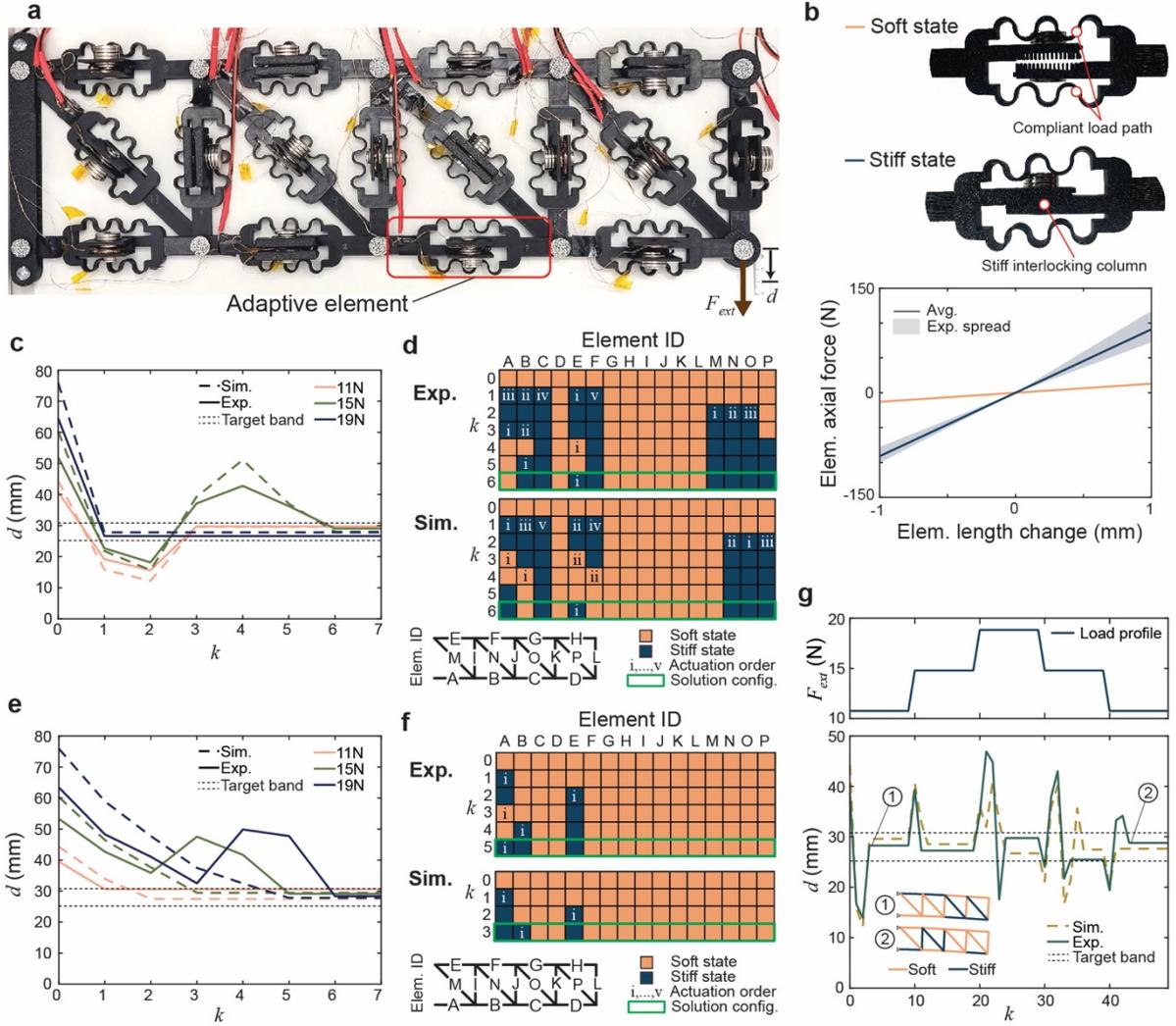

**Fig. 3 | Experimental demonstration of autonomous real-time adaptation via physical learning in a metamaterial beam. a.** Experimental implementation of metamaterial cantilever beam, consisting of 16 adaptive elements (see Methods section 'Reprogrammable metamaterial testbed' for details on the adaptive element). **b.** (top) Different load paths in the adaptive element. In the soft state (13 N/mm), the load path is through compliant circular arches. In the stiff state (91 N/mm), the load path is through a stiff central column with interlocking teeth. (bottom) Uniaxial test data for an adaptive element for its 2 stiffness states. There is experimental spread in the soft state, but it is not visible in this plot. **c.** Tip displacement response of the metamaterial beam with physical learning applied. Three external loads are tested with the hyper-parameters $p_0 = 0.31$, $\gamma = 0.7$. **d.** Sequence of actuations corresponding to $F_{ext} = 15$ N in Fig. 3c, in experiment (exp.) and simulation (sim.). The roman numerals in the cell indicate that the element was actuated in that iteration and denote the strain deviation magnitudes in descending order (and the actuation order). **e.** Tip displacement response with physical learning applied. Three external loads are tested using a constant one actuation per iteration, i.e. $n_c = 1$ ($p_0 = 0.06$, $\gamma = 1.0$). **f.** Sequence of actuations corresponding to $F_{ext} = 15$ N in Fig. 3f. **g.** Tip displacement response (bottom) under a time-varying load profile (top). Experimental stiffness configuration solutions for the same external load ($F_{ext}=11$ N) at two different points in the adaptation history are shown.

We repeat the adaptation experiment with $p_0 = 0.06$, $\gamma = 1.0$, i.e., constant $n_c = 1$ for all $k$. These are not Pareto-optimal parameters. In this case, the experimental and simulated adaptation actions and tip displacements diverge (Fig. 3e), as compared with the Pareto-optimal hyper-parameters. This discrepancy arises because when there is only one actuation per iteration, any difference in actuation dominates the outcome, leading to very different pathways to the solution. With multiple actuations, the effect of a single variation is diluted . We note that the experiment finds the same stiffness solution as the simulation but takes more iterations (Fig. 3f). As such, it is more difficult to predict experimental performance for sub-optimal parameters. Compared with the Pareto-optimal parameters $p_0 = 0.31$, $\gamma = 0.7$, the accuracy is similar ($\bar{e}_s$=4.9%), but the learning rate is slower ($\bar{k}_s$=4.0). Regardless, we see that adaptation remains successful even though the tip displacement and actuation history varied, further highlighting the advantage of leveraging mechanics for computing.

We further examine physical learning under a time-varying load, with a step change in $F_{ext}$ every 10 iterations (Fig. 3g). Note that this is a quasi-static problem, given that adaptation completes prior to each load change. In this case, physical learning searches for a new stiffness configuration whenever the tip displacement exceeds the target band, and the metamaterial beam is able to continually adapt (Fig. 3g). Notably, whenever $F_{ext}$ changes, the metamaterial starts in the stiffness configuration from the previous load interval, highlighting that successful adaptation is independent of the initial stiffness configuration. Interestingly, the adapted stiffness distribution can differ depending on the initial configuration. For instance, for $k = 0$-$9$ and $k = 40$-$49$, $F_{ext}$ is the same at 11 N, but the adapted tip displacements differ (28.3 mm and 28.8 mm, respectively), with different stiffness configuration solutions for both cases (Fig. 3g). This suggests that the intrinsic behavior of the algorithm is to find locally optimum solutions.

**Robust adaptation in damaged metamaterials**

Our physical learning approach exhibits tolerance to damage. Here, a damaged element is defined as one that loses sensing and reprogramming ability, and is stuck in the soft state. We repeat the experiment but damage one adaptive element that was nominally actuated in the undamaged adaptation. The hyper-parameters $p_0 = 0.31$, $\gamma = 0.7$ are used here. There are no changes to the model-free algorithm, and the element with the next highest strain deviation actuates in place of the damaged element. We observe that adaptation remains successful across the test cases $F_{ext} \in \{11,15,19\}$(N) (Fig. 4a), yielding $\bar{k}_s = 2.7$ and $\bar{e}_s = 1.0\%$. This damage tolerance characteristic is attributed to the physical learning being model-free and the high number of available DOFs providing high redundancy.

For a comprehensive evaluation, we repeat the damaged adaptation in simulation for each element for 9 load cases from 11N to 43N. We find that adaptation rate remains very high across cases where convergence remains physically possible despite damage. Here, we report an in-feasibility-domain adaptation rate (success rate over cases where the load is within the structure's reduced operating range) of 96.3%. Also, the performance of the damaged metamaterial ($\bar{k}_s = 3.7$, $\bar{e}_s = 3.1\%$) is similar to the undamaged system ($\bar{k}_s = 4.1$, $\bar{e}_s = 2.2\%$). On the other hand, the overall adaptation rate (success rate over all damage cases) is 81.3%, slightly lower but still relatively high. Here, some cases fail to adapt when the loss of critical DOFs, such as elements near the cantilever root, causes the applied force to exceed the structure's reduced capacity. For example, the overall adaptation rate is the lowest when the bottom-left element near the root is damaged, increasing as the damaged element moves closer to the tip (Fig. 4b). In contrast, the in-feasibility-domain adaptation rate does not depend on the location of the adaptive element (Fig. 4b).

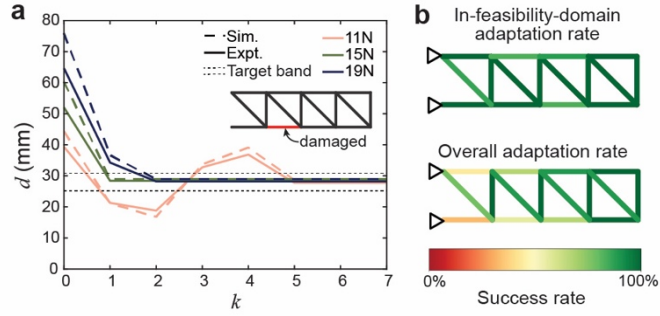

**Fig. 4 | Robust adaptation via physical learning in metamaterials with damaged elements. a.** Tip displacement response of the metamaterial beam when applying physical learning ($p_0 = 0.31$, $\gamma = 0.7$), with one damaged adaptive element as indicated (inset). **b.** Successful adaptation rate in simulation measured as a function of the damaged element location. The in-feasibility-domain adaptation rate considers a subset of damaged cases where adaptation is still physically possible, whereas the overall adaptation rate considers all damaged cases.

**Broader applications**

Physical learning is applicable across lattice architectures, owing to its model-free characteristic. As an illustration, we use a bending-dominated metamaterial beam, where the deformation involves shear and non-uniform bending of each element (Fig. 5a). Each element has two variable bending stiffness states (Ext. Data. Fig. 4a), and can be implemented physically using reversible lamination via electrostatic jamming and dry adhesives as in Chen et al.[21]. Twelve hinges are arranged into an anti-tetrachiral lattice[41] (Fig. 5a). The mechanical response of the structure is simulated in finite element analysis using experimentally informed test data (see Methods section 'Bending-dominated lattice'). We impose the same tip displacement adaptation task as in Eq. 1, with $F_{ext} \in \{0.3,…,1.3\}$ (N), $d_t = 27$ mm, $\delta = 10\%$, and apply the model-free algorithm with no modifications. Using the Pareto-optimal $p_0 = 0.25$, $\gamma = 0.9$, we achieve $\bar{k}_s = 4.1$ and $\bar{e}_s = 3.6\%$ (Fig. 5a). Physical learning shows rapid adaptation with high accuracy across lattice geometries, even for this highly non-linear bending-dominated lattice. One advantage of the bending-dominated lattice is potential use in larger deformation applications.

Physical learning is similarly successful for complex target shapes and distributed loading. We consider a reprogrammable metamaterial airfoil (Fig. 5b) composed of a lattice of 65 adaptive elements, obtained by discretizing a NACA 2412 airfoil (see Methods section 'Shape adaptation of airfoil'). We task the airfoil with maintaining the target shape of the outer surface under an unknown distributed load. Here, the arbitrary target shape is derived from a small constant load of 1N applied to the top surface nodes. We apply three different distributed load profiles (Fig. 5b) to cause the airfoil shape to deviate from its target, resembling random wind gusts during cruising flight. The airfoil recovers its target shape through physical learning, with the stiffness configuration solutions for the three loading cases shown in Fig. 5c. With $\delta = 2.3\%$ (normalized against the maximum airfoil thickness), $p_0 = 0.08$, $\gamma = 0.9$, we achieve $\bar{k}_s = 5.0$ and $\bar{e}_s = 1.1\%$ (root-mean-square-error across all nodes). Here, the target shape can be viewed as a morphed aircraft wing, while the load profiles represent unknown external stimuli such as wind gusts during cruising flight. While the exact magnitudes and shapes here are not representative of specific application requirements, the ability of our physical learning approach to simultaneously minimize the displacement error across multiple nodes demonstrates its applicability to complex, multi-objective shape adaptation problems. Potential applications include optimal airfoil morphing for each set of conditions within a flight envelope.

Finally, we also explore the applicability of physical learning to the dual problem – force control. Using an axially-dominated metamaterial beam with $n = 16$, $m = 2$, we consider maintaining a tip reaction force, $F$, under an unknown tip displacement, $d_{ext}$ such that $|F(\mathbf{K})-F_t| < \delta$, where $F_t$ is the target tip force (Fig. 5d). This setup resembles a robotic finger optimizing contact to an object with unknown shape. We invert the stress-strain (and force-displacement) constitutive relations that describe the structure's mechanics and use target stresses for physical learning instead of target strains (see Methods section 'Force control'). For $F_t = 15$N, $d_{ext} \in \{15, 20, 30\}$ (mm), $\delta = 5\%$, we achieve $\bar{k}_s = 3.3$ and $\bar{e}_s = 2.2\%$, with the desired reaction force successfully achieved in all cases (Fig. 5d). Given that force control is commonly employed in many grasp optimization strategies[42], we envision potential applications in robotic grippers for optimizing gripping of unknown objects.

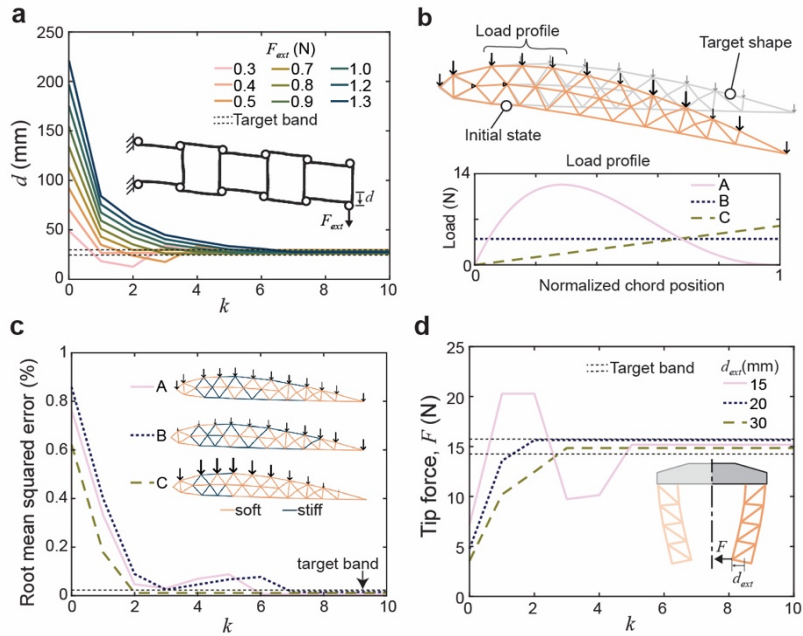

**Fig. 5 | Broader applications of physical learning. a.** Physical learning in bending-dominated metamaterial with 12 elements in an anti-tetrachiral structure. Tip displacement response as a function of iteration across various load cases is shown. **b.** (top) Metamaterial airfoil is tasked with adapting towards a target shape (collection of outer surface nodal displacements) under unknown distributed load profiles applied to the top surface. (top) Three different load profiles studied. **c.** Root mean squared error response of outer surface displacements, normalized against the maximum airfoil thickness, with target accuracy of $\delta = 2.3\%$. (inset) Stiffness configuration solutions for each of the three cases. The cases A, B, C correspond to the load profiles in Fig. 5b. **d.** Reaction force, $F$, response of metamaterial beam by applying physical learning ($p_0 = 0.31$, $\gamma = 0.7$), under three cases of unknown externally applied tip displacement, $d_{ext}$.

## Discussion and outlook

This study presents a physical learning approach for real-time adaptation control of mechanical metamaterials with discrete adaptive stiffness. We leverage structural mechanics for computing to embody iterative assessment and learning functionalities. Given a change in environmental conditions, physical learning decides on the appropriate actuation actions to achieve a target

response – a non-convex, ill-posed, and discrete inverse problem, typically solved via non-convex optimization. Physical learning enables autonomous operation of adaptive structures in unknown environments without pre-computing adaptation strategies and instead leveraging mechanical intelligence.

Physical learning offers several advantages owing to its model-free nature. First, the learning rate is very fast (<10 iterations) and remains fast even in structures with hundreds of DOFs, all while accurately achieving target behaviors (~1%). Combined with adaptive elements that can actuate on the order of ~100 ms, our approach achieves real-time adaptation on the order of seconds. This rapid adaptation is crucial to the control of next-generation reprogrammable structures, which have complex architectures with increasingly growing number of DOFs that make model-based controls challenging. Second, physical learning is robust to manufacturing imperfections, sensor noise, and damage to the metamaterial structure. This robustness is pivotal to operation in extreme, inaccessible environments such as in deep space and underwater exploration where it would be impractical to repair damaged structures. Finally, physical learning scales to various mechanical metamaterial architectures, complex shape control, and even force control.

The model-free nature of physical learning opens the door for further embodying the sense-assess-respond functionalities directly into adaptive structures. The remaining rudimentary electronics in the current approach could be reduced or eliminated in the future using mechanical digital logic gates[32,43] and analog mechanical computing[44,45] to carry out sensing and algebraic operations in the model-free algorithm. We believe that the simple logic in physical learning could pave the way towards an inherently intelligent mechanical structure in which the sensing, computation, and actuation function are fully integrated and purely mechanical. This offers potential applications in harsh space environments where ionizing radiation and extreme temperatures can rapidly degrade electronics[46], or in encryption by leveraging purely mechanical functionality and preventing electromagnetic attacks.

**Code availability.** The codes that support the findings of this study are available at: (link will be made available).

**Acknowledgements.** This research is sponsored by the Defense Advanced Research Project Agency (DARPA) through the Young Faculty Award (#D24AP00327). The content does not necessarily reflect the position or policy of the government, and no official endorsement should be inferred.

**Author contributions.** K.J.C. and M.S. conceived the research problem; C.C., C.S., J.M. and E.A. designed and fabricated the experimental setup and conducted experiments; K.J.C., E.C. and A.M. performed analysis; K.J.C., E.A. and M.S. wrote the manuscript. All authors revised the manuscript.

**Competing interests.** The authors declare no competing interests.

## Methods

## Model-free algorithm

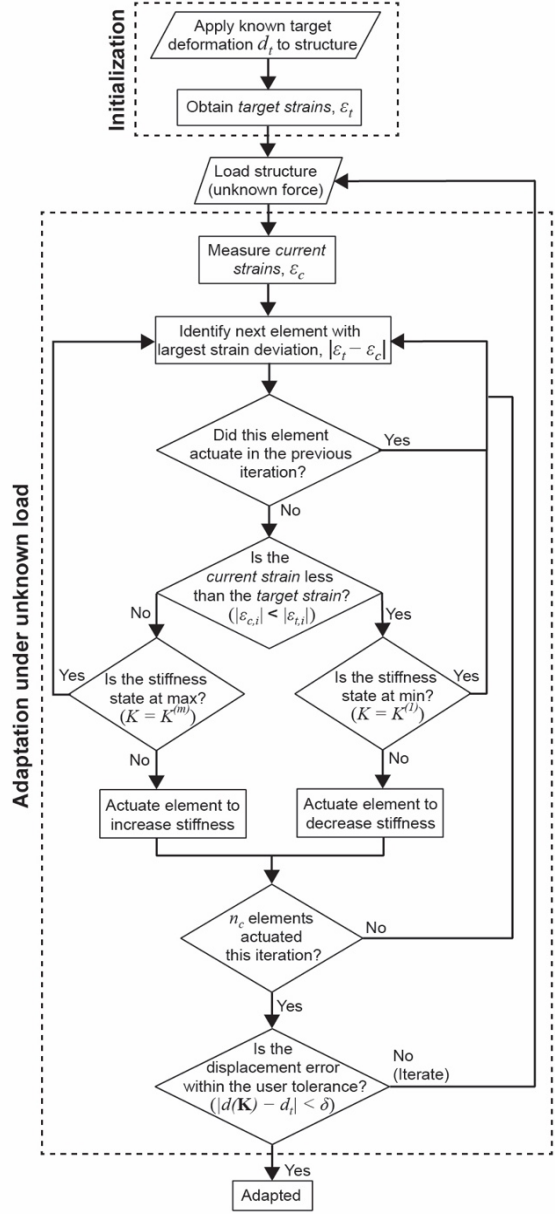

**Extended Data Fig. 1 | Detailed flow-chart of the model-free algorithm in our physical learning approach.**

A flow-chart of the model-free algorithm described in the main text is shown in Extended Data Fig. 1. We find that introducing short-term memory greatly increased the successful adaptation rate for all hyper-parameters tested (Extended Data Fig. 2a). With $\delta = 10\%$, the number of successful hyper-parameters increased sevenfold from 4.6% to 34%. Short-term memory helps overcome local minima in the rugged non-convex solution space by preventing repeated actuations of the same adaptive element across iterations. We limit the short-term memory to only one iteration to minimize the effect on learning rate.

We terminate the physical learning process once the tip displacement error relative to the target $|(d-d_t)/d_t|$ reaches below the user-prescribed error tolerance, $\delta$. This was implemented to 'fix' the stiffness configuration of the metamaterial structure in the solution state. Physical learning does not evaluate the tip displacement, which is the global objective function. Without the termination check, the algorithm would continue to actuate elements, resulting in a response where the tip displacement oscillates about the target (Extended Data Fig. 2b). For the $n = 16$, $m = 2$ metamaterial beam ($F_{ext} = 15$ N, $d_t = 28$ mm, $p_0 = 0.31$, $\gamma = 0.7$), these oscillations are quite large and could cause the tip displacement to oscillate outside the target band. As such, this termination check, which is already used in most conventional optimization algorithms, is essential in this case. However, for a metamaterial with more DOFs, e.g., $n = 16$, $m = 20$ or $n = 1240$, $m = 2$ (with hyper-parameters adjusted accordingly), the magnitude of these oscillations decreases substantially (Extended Data Fig. 2b). In the steady state, the tip displacement response stays entirely within the target band. Intrinsically, the physical learning approach causes the tip displacement to approach the target, while the number of DOFs affects the magnitude of oscillations. (Refer to Main Text section 'Characterizing learning rates and accuracy' for detailed study on $m$ and $n$). Hence, with many DOFs, the termination check is no longer necessary. This can be advantageous because it is difficult to measure the shape in situ, which is required to evaluate the global objective function. Another advantage is the reduced computational complexity.

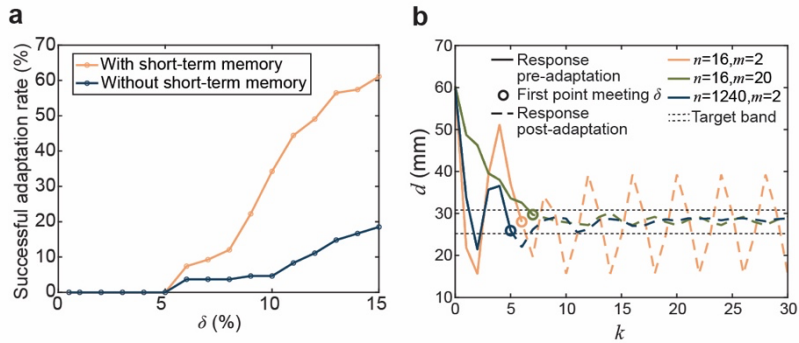

**Extended Data Fig. 2 | Supporting data for derivation of proposed model-free algorithm. a.** Successful adaptation rate over all hyper-parameter and load cases, plotted against the convergence target, $\delta$. The $n = 16$, $m = 2$ metamaterial beam is used here. **b.** Tip displacement response (for $F_{ext} = 15$ N) pre- and post-adaptation check for metamaterial beam for varying $n$ and $m$. The dot indicates the point at which physical learning stops with the termination check. The dashed lines indicate the tip displacement response without the termination check.

### Simulation of metamaterial beam

To simulate the mechanical response of the axially-dominated metamaterial structures, we implement a finite element (FE) model of the cantilever beam lattice in Python based on the non-linear matrix stiffness method for 2D truss structures[1]. The constitutive equation is:

$$\bm{F} = (\bm{T}^T(\bm{K_M} + \bm{K_G})\bm{T})\bm{d}, \qquad (4)$$

where $\bm{F}$ and $\bm{d}$ are 2D force and displacement vectors of all nodes in the truss, respectively, $\bm{T}$ is the transformation matrix from the truss element frame to global frame, $\bm{K_M}$ is the elastic stiffness matrix dependent on material and truss properties, and $\bm{K_G}$ is the geometric stiffness matrix, included to account for geometric non-linearity and large deformations. To reduce computational complexity, the adaptive elements are represented in simulation as linear-elastic 1D truss elements. Stiffness values are assigned depending on whether the adaptive element is in the soft (13 N/mm) or stiff state (91 N/mm), as informed by experimental data (Main Text Fig. 3b). The lattice model is pin-jointed and pinned at the root. A point load is applied at the tip. Nodal displacements of adaptive elements are tracked in a fixed global coordinate system. Finally, since $K_G$ is a function of $d$, an iterative Newton-Raphson scheme is used to solve for the deformation, with the global stiffness matrix updated at each iteration. The solver terminates once the residual norm, $\|F - (T^T(K_M+K_G)T)d\|$, falls below $10^{-6}$. In the post-processing step, axial strains are extracted from the deformed truss elements and passed to physical learning, which returns the indices of adaptive elements that require stiffness updates in the next iteration. The same methodology is used to simulate the airfoil lattice.

**Parametric study of lattice parameters**

In simulation, we independently examine the effect of varying the number of adaptive elements, $n$, and the number of stiffness states in each element, $m$. First, we vary $n \in \{16, 56, 208, 616, 2140\}$, while keeping $m = 2$ constant. These values are chosen to keep the ratio of horizontal to vertical elements constant, representing lattices with 1×4, 2×8, 4×16, 7×28, 10×40 elements, respectively. Second, while keeping $n = 16$ constant, we vary $m \in \{2, 4, 10, 20, 40\}$ while maintaining the minimum and maximum stiffness states $K^{(1)}$ and $K^{(m)}$ at 13 and 91 (N/mm), respectively. For a given $m$, intermediate stiffness values are equally distributed, i.e. $K^{(j+1)} - K^{(j)} = K^{(j)} - K^{(j-1)}$. Here, $\delta$ was varied from 0.02% to 20% and the hyper-parameters are varied with $(p_0, \gamma) \in [0.02, 0.94] \times [0.1, 1.0]$ to generate the Pareto fronts.

**Reprogrammable metamaterial testbed**

*Adaptive elements*

The adaptive elements are additively manufactured using a Bambu Lab X1C 3D printer using acrylonitrile butadiene styrene (ABS) (Extended Data Fig. 3a). The specific element design is selected to achieve equal and linear response in both tension and compression. In the stiff configuration, the central columns are interlocked via teeth, enabling the transfer of axial loads. A latching mechanism using C-clips secures the interlocked state. In the soft configuration, the columns are separated, and axial loads are supported by compliant ABS springs.

To characterize the stiffness of the adaptive elements, uniaxial tensile and compressive tests are performed on sixteen specimens using an Instron 68TM-50 universal testing machine equipped with a 2530-100N load cell. The results exhibit low variability across samples and demonstrate a linear stiffness response in two stiffness states (mean $R^2$=0.99) (Main Text Fig. 3b). The soft state exhibits an average stiffness of 13 N/mm, while the stiff state averages 91 N/mm, corresponding to a stiffness ratio of 7.

Transition between stiffness states is controlled via an embedded actuation system combining electromagnets and permanent magnets. The goal of the actuation system is to engage or disengage the C-clip latching mechanism, which maintains the element in the stiff configuration by keeping the interlocking teeth in contact. Each electromagnet consists of 150 turns of 30 AWG polyurethane-enamelled copper wire (Extended Data Fig. 3a), wound around a cylindrical form with an inner diameter of 8 mm and an outer diameter of 22 mm. By reversing the current direction, the solenoids can produce both attractive and repulsive forces. They are bonded to the ABS panel walls of each element using epoxy. Neodymium permanent magnets are positioned on opposing panels. We note that the C-clips used to hold the stiff state are susceptible to fatigue after a large number of cycles.

*Adaptive metamaterial testbed*

We fabricate a 1×4 cell metamaterial cantilever beam (Main Text Fig. 3a) composed of the adaptive elements ($n = 16, m = 2$). These elements are assembled using long screws to approximate pin-jointed connections within the lattice. To minimize friction during testing, 3D-printed supports are placed in contact with a low friction PTFE sheet. The root of the beam is pinned to an optical table using screws.

Strain is measured using 350 Ω high-precision resistive strain gauges, bonded with cyanoacrylate adhesive to the outer surface of the central wave region of the ABS springs. Each gauge is configured in a quarter-bridge circuit and interfaced with an HX711 load cell amplifier, providing a gain of 64 and 24-bit analog-to-digital conversion (Extended Data Fig. 3b). To convert raw signals into global element strain values, four calibration factors are employed, accounting for the two stiffness states (stiff and soft) and two loading modes (tension and compression). Calibration factors are obtained through tensile and compressive testing on the Instron load frame, with Digital Image Correlation (DIC) used for calibration. The calibration factors convert local strain on the ABS springs to a global element length change. The two quantities exhibit a linear correlation. Data acquisition is handled by an Arduino MEGA microcontroller, with all HX711 modules synchronized via a shared clock line.

To apply tip loading to the metamaterial beam, an inextensible cable is connected to a suspended dead weight. The load (Extended Data Fig. 3c) consists of a slotted weight set in a custom 3D-printed holder, with small steel marbles added for fine weight adjustment. The weight is reversibly applied and offloaded via a rack and pinion mechanism using a NEMA 17 stepper motor (59 N·cm, 2 A/phase) via a gear transmission, and controlled using a DRV8825 stepper driver.

The electromagnets are driven with 30 V and 5 A pulses for 40 ms and 100 ms to transition into the soft and stiff states, respectively, using a Tekpower TP3005T DC regulated power supply. Each electromagnet is controlled via a dedicated DRV8873 H-bridge driver (Extended Data Fig. 3b), with switching signals provided by an Arduino MEGA microcontroller through two MCP23017 I/O expanders (16 GPIOs each, over I²C). The microcontroller coordinates all system components, ensuring synchronized actuation, sensing, and loading operations.

The model-free algorithm is implemented on an Arduino using C++. Strain measurements serve as input to the controller, which rank the elements based on strain deviations from targets, select those to actuate, and send the corresponding commands to the I/O expanders to drive the solenoids.

*Metamaterial Deformation Validation*

Digital Image Correlation (DIC) is used as a secondary check to validate convergence to the target tip displacement. DIC data is not used for physical learning, as tip displacement is not used in actuation decisions. The DIC system uses two Basler boost boA4112-68 cameras equipped with Schneider JADE 2.8/25 C lenses set to an f/8.0 aperture. Speckled targets are affixed to the lattice nodes to serve as high-contrast tracking markers. Image acquisition is performed using Vic-Snap 9, and subsequent DIC analysis is conducted with Vic-3D software (Correlated Solutions). System calibration is carried out using a 14×10 standard calibration target with 28 mm dot spacing. Final data processing and visualization are performed in MATLAB.

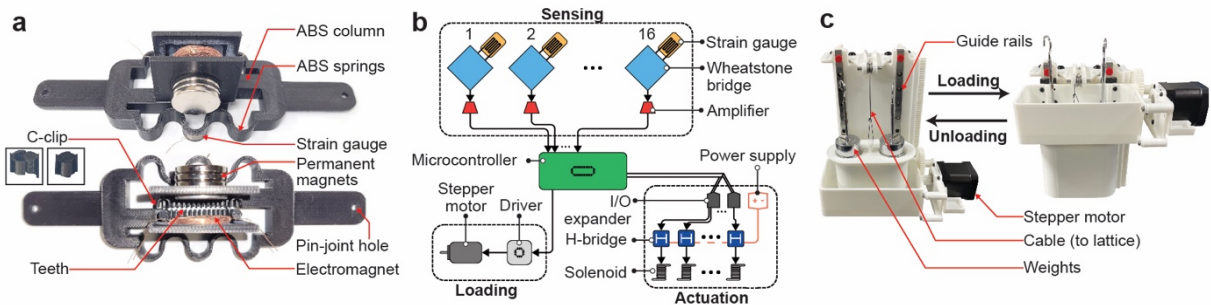

**Extended Data Fig 3. | Constituents of the physical learning testbed. a.** 3D-printed adaptive elements. **b.** Electronics schematic illustrating sensing, loading, and actuation. **c.** Photographs of the loading mechanism, showing the cable under tension (left) and slack (right).

**Bending-dominated lattice**

Hinges with adaptive bending stiffness (Extended Data Fig. 4a) have been previously introduced by the authors[2,3]. From experimental testing, they have two available stiffness states of 0.44 N/mm and 4.06 N/mm, giving a ratio of 9.15, and can be assembled into an anti-tetrachiral metamaterial lattice. An FE model (Main Text Fig. 5a inset) of the bending-dominated cantilever beam is implemented using the commercial software Abaqus/Standard[4]. We use a homogenized representation of the adaptive hinges as a continuum beam to minimize computational cost, assigning each hinge an effective elastic modulus based on its current stiffness state. Each hinge is meshed using 1D linear beam elements (B21) and assigned 10 elements along its length, as determined through a mesh convergence study. Next, we assemble the lattice based on the structural constraints of the anti-tetrachiral lattice geometry, clamping hinge ends to their corresponding reference point node. Clamped boundary constraints are assigned to the root and a point load is applied to the tip. Finally, a geometrically non-linear static analysis solves for the mechanical deformation response. From the FE analysis results, strain readings are extracted from the outer surface of the hinges' center. Finally, strain readings are converted into absolute values and averaged over the top and bottom surfaces, since the bending direction is not important.

We systematically explored the parameter space $(p_0, \gamma, \delta) \in [0.08, 0.92] \times [0.1, 1.0] \times [0.07, 0.15]$ to generate a Pareto front (Extended Data Fig. 4b), and identified $p_0 = 0.25$, $\gamma = 0.9$ as one Pareto-optimal hyper-parameter set.

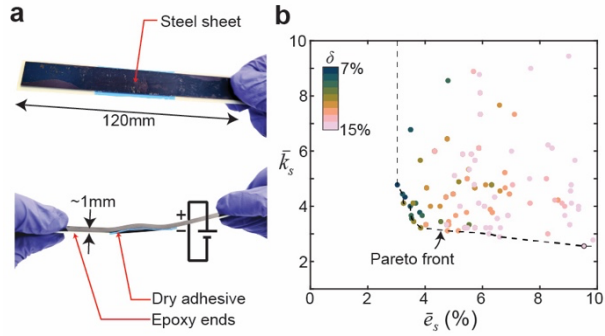

**Extended Data Fig 4. Supporting data for the bending-dominated metamaterial beam | a.** Photos of the bending-adaptive element. **b.** Pareto front for the bending-dominated lattice metamaterial beam with $n = 12$, $m = 2$.

## Shape adaptation of airfoil

The metamaterial airfoil adaptation task is studied in simulation. We construct a metamaterial airfoil consisting of $n = 65$ adaptive elements by discretizing a NACA 2412[5] airfoil using the open-source finite element mesh generator Gmsh[6] and converting edges to adaptive elements. We scale the airfoil dimensions to achieve a mean adaptive element length of 140 mm (to match experiments) with a standard deviation of 24.8%. The scaled chord length is 1800 mm. Adaptive elements are assigned the same $m = 2$ available stiffness states as in the experimental setup. The mechanical response of the structure is simulated using the same finite-element code as for the metamaterial beam (Methods section 'Simulation of metamaterial beam'). For the airfoil shape adaptation task, we modify the tip displacement task from equation (1) as follows:

Find $\boldsymbol{K} = \{K_1, K_2, \ldots, K_{65}\}$ where $K_i \in \{13, 91\}$ (N/mm) s. t. $\text{RMSE}(\boldsymbol{d}(\boldsymbol{K}) - \boldsymbol{d_t})/h < \delta$, (5)

where $\boldsymbol{d}$ is the nodal displacement vector of the nodes of interest and is a function of $\boldsymbol{K}$, $\boldsymbol{d_t}$ is the target nodal displacement vector of the nodes of interest, RMSE is the root-mean-square-error between $\boldsymbol{d}$ and $\boldsymbol{d_t}$ and is normalized against the maximum airfoil thickness $h$ (216 mm), and $\delta$ is a user-specified tolerance parameter. The nodes of interest are those along the outer surface of the metamaterial airfoil.

We consider a distributed load profile applied over the top surface. Since the metamaterial consists of discrete elements, we convert the distributed load profile into discretized loads applied at the nodes along the outer surface. The target shape, $\boldsymbol{d_t}$, and target strains are obtained by applying a constant 1 N to each node along the top surface. Three arbitrary load profiles (Main Text Fig. 5b) are considered as external disturbances: Profile A is a constant 4N, Profile B is linearly increasing from 0 N at the leading edge to 6 N at the trailing edge, and Profile C follows a beta distribution with the discretized load proportional to $(x/c)(1-x/c)^3$, where $x/c$ is the normalized coordinate in the chordwise direction, and scaled to give a max point load of 12 N. To prevent rigid body motion of the metamaterial airfoil, two interior nodes are pinned near the quarter-chord position, as indicated by the two nodes with a triangle (Main Text Fig. 5b).

## Force control

The force control task is formulated by modifying Equation (1) as follows:

Find $\boldsymbol{C} = \{C_1, C_2, \dots, C_{16}\}$ where $C_i \in \left\{\frac{1}{91}, \frac{1}{13}\right\}$ (mm/N) s.t. $|(F(\boldsymbol{C}) - F_t)/F_t| < \delta$, (6)

where $C = 1/K$ is the compliance. In the model-free algorithm, we use stresses, $\sigma$, and compliance, $C$, as analogues for strains, $\varepsilon$, and stiffness, $K$, respectively. Stresses are derived by multiplying the measured strain by the effective modulus at the element's current stiffness state. From here, we can implement the algorithm as before. The target stresses, $\sigma_{t,i}$, are derived from applying the target tip force in an arbitrary stiffness configuration. We initialize all elements to the highest compliance state. Then, each adaptive element updates its compliance state, $C_i$, following the principle of locally minimizing stress deviations, $\sigma_{d,i} = |\sigma_{c,i} - \sigma_{t,i}|$. If the current stress exceeds magnitude exceeds its target, i.e., $|\sigma_{c,i}| > |\sigma_{t,i}|$, then the element increases its compliance to the next discrete state, except if it is already at the maximum compliance. Conversely, the element is softened if $|\sigma_{c,i}| < |\sigma_{t,i}|$ and it is not already in the minimum stiffness state. Finally, we regulate the number of actuations and implement the short-term memory as before.

**Supplementary 1 – Benchmark against conventional algorithms**

To establish a general performance benchmark for the model-free algorithm in our physical learning approach, we compare its accuracy and learning rate with that of conventional optimization algorithms. Here, we consider the tip displacement control of the extension dominated cantilever beam problem (Main Text Eq. 1). This problem involves truss elements with discrete stiffness states, so the algorithm must be able to handle discrete inputs. Furthermore, gradient information is unavailable and so gradient-based optimization methods are not suitable for this problem. Accounting for these restrictions, we select two off-the-shelf optimization algorithms to represent a diverse set of strategies: Partial Pattern Search (PPS) — a direct search-based method, and Genetic Algorithm (GA) — a population-based method. It must be noted that the logistics in physically implementing these conventional algorithms differ from the model-free algorithm. Specifically, the tip displacement would be measured instead of element strain readings, and the increased computational requirements of the conventional optimization algorithms necessitate incorporating more powerful computers. Therefore, these algorithms are not intended for direct comparison. Instead, this evaluation is meant to offer an overall sense of the performance of the physical learning algorithm in terms of accuracy and learning rate. In the subsequent paragraphs, we describe the methodology of the three algorithms.

PPS iteratively explores the search space using a greedy strategy, evaluating candidate solutions along search directions corresponding to randomly selected individual truss actuations[1]. In the context of the cantilever lattice problem, the search space comprises all possible permutations of stiffness configurations of the truss elements. Initially, all truss elements are assigned the soft stiffness state, and the tip displacement error is evaluated. Subsequently, a single truss element is actuated at random. If this stiffness actuation reduces the tip displacement error, then the actuation is accepted. Otherwise, the actuation is rejected, and another truss element is actuated at random. This reject-and-retry process continues until either all possible actuations for the current configuration is exhausted, in which case the current stiffness configuration is accepted as the final solution, or a stiffness actuation reduces the tip displacement error and is accepted. The search process iterates until a stiffness configuration with tip displacement error within error tolerance is identified and accepted as the final solution. Given the stochastic nature of PPS, the algorithm was repeated 50 times for each load case to characterize average performance.

GA explores the search space by evolving a population of candidate solutions through selection, crossover, and mutation operations[2]. For the cantilever lattice problem, a population of 20 random stiffness configurations (samples) is initialized. Then, each sample is evaluated to obtain the tip displacement error. If the best performing sample meets the error tolerance parameter, then that sample is accepted as the final solution and the algorithm terminates. Else, 3 elite samples with the smallest errors are selected for crossover to generate a child. The child is then mutated with a 10% probability of each truss element actuating to generate 19 samples and combined with the best sample from the parent generation to create a new generation of 20 samples. The new population is evaluated again, and the process iterates until the error tolerance is met. Here, we consider each evaluation of a candidate solution as one iteration, which means that each generation in the GA consists of 20 iterations. Just like the PPS, GA is stochastic, and so the algorithm is repeated 50 times for each load case to characterize average performance.

We implement PPS and GA on the same displacement control problem for the $n = 16$ and $n = 56$ metamaterial beams, with $m = 2$ in both cases. A tolerance parameter $\delta$ from 1% to 15% is used to generate Pareto fronts for PPS and GA. Pareto fronts for PPS, GA and physical learning representing mean and worst-case performance are shown in Supplementary 1 Fig. 1a and 1b, respectively. Given the stochasticity of PPS and GA, these algorithms sometimes fail to converge. Therefore, only data points achieving >90% convergence rates over the 50 repeats are used.

Physical learning exhibits a faster learning rate, requiring fewer iterations to achieve a given mean error value. Notability, when required accuracy is moderate (>3% for $n = 16$ and >1% for $n = 56$), physical learning requires significantly fewer iterations on average to find a solution. In contrast, conventional optimizations approaches can achieve lower errors as physical learning finds local minima.

Conventional optimization requires substantially more iterations to achieve a given error value for structures with a higher number of degrees of freedom (DOF), revealing a scalability problem. The same was not observed for physical learning, which maintained comparable performance across a different number of DOFs. Finally, the stochastic nature of PPS and GA means that the worst case performance (Supplementary 1 Fig. 1b) can be an order of magnitude worse that the average performance. Meanwhile, worst-case performance of physical learning is only a handful of iterations higher than average performance.

Given the rapid learning rate, the deterministic nature, and scalability of physical learning, it is suited to applications where there are time-varying disturbances and consistent rapid adaptation is required.

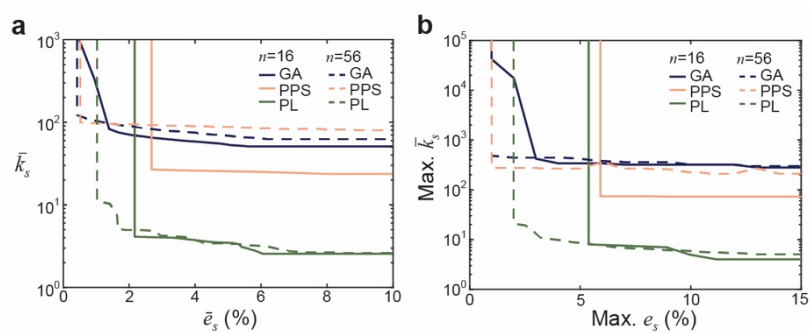

**Supplementary 1 Fig. 1 | Physical learning compared against partial pattern search and genetic algorithm. a.** Pareto fronts of mean performance values for the $n = 16$ and $n = 56$ structures for physical learning (PL), partial pattern search (PPS) and genetic algorithm (GA). **b.** Pareto fronts of worst case performance values for the $n = 16$ and $n = 56$ structures for PL, PPS and GA.

**Supplementary 4 – Characterization of sensor noise**

Physical learning uses the strain of adaptive elements to make actuation decisions. In the finite element (FE) simulation, these strain readings are noise-free. In practice, strain gauges are subject to measurement noise. Unlike manufacturing imperfections and friction, sensor noise cannot be inherently captured in the mechanics of the metamaterial. It is therefore important to assess how such variability might affect both the speed and accuracy of adaptation by physical learning. In this supplementary section, we characterize the noise distribution present in the experimental strain measurements and quantify the impact of this distribution on the convergence behavior.

*Deriving the noise distribution*

Measurement noise in strain gauges originates from various stochastic sources, including thermal agitation, amplifier noise, and circuit interference. By the central limit theorem, the sum of these independent effects is expected to follow a Gaussian distribution. Since such noise is centered around zero, only the standard deviation needs to be determined.

We consider three load cases: 1) Unloaded metamaterial beam, 2) Metamaterial beam loaded with 286 g, 3) Metamaterial beam loaded with 486 g. In each case, 500 strain measurements are recorded from each strain gauge. In these measurements, we observe a gradual drift in the strain reading over a 25 minute acquisition period, likely due to thermal effects, which we correct for using a first-order fit. Then, we fit the drift-corrected data to a Gaussian, and evaluate the fit quality via the Kolmogorov–Smirnov (KS) statistic. Here, only datasets with KS ≤ 0.1 are retained for subsequent analysis, since higher KS values indicate a departure from Gaussian noise and would render the standard deviation non-representative. Higher loads are associated with lower noise standard deviations.

For the noise sensitivity analysis that follows, we take a conservative approach and use the maximum measured standard deviation across all cases. We therefore use the following noise distribution:

$$\varepsilon \sim \mathcal{N}(0, \sigma^2), \quad \sigma = 2.93 \times 10^{-5} \tag{S4-1}$$

*Noise sensitivity of physical learning*

We artificially introduce this noise distribution to the strains measured from the FE simulation additively. At each iteration, noise is sampled independently for each gauge from the measured distribution, so that gauges have different noise levels at any given iteration. Under this noisy condition, we quantify the learning performance using the same metrics as in the main text, i.e. learning rate and accuracy. Here, we consider just one load case, $F_{ext}$ = 15 N, and one error tolerance, $\delta$ = 10%. A Monte Carlo approach is used, with 1000 runs per noise level. The mean and standard deviation of each metric are computed across runs, yielding the results shown in Supplementary 4 Fig. 1a and 1b. We also plot the result corresponding to the measured noise level in the experimental setup (indicated by the beige dot).

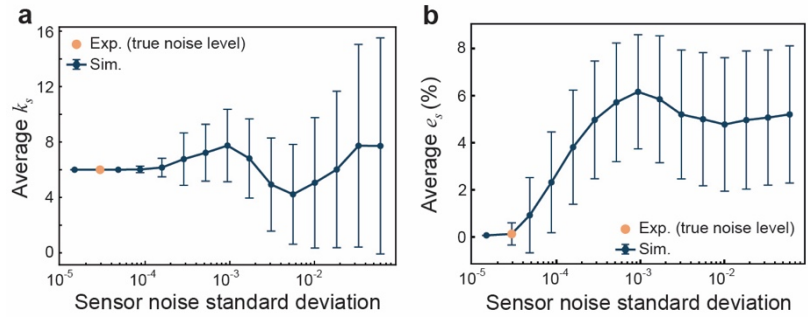

**Supplementary 4 Fig. 1 | Effect of increasing sensor noise on the adaptation performance of physical learning. a.** Learning rate versus the sensor noise standard deviation. **b.** Tip displacement error versus the sensor noise standard deviation. For both plots, the parameters used are $n = 16$, $m = 2$, $d_t = 28$ mm, $F_{ext} = 15$ N, $\delta = 10\%$. The dark blue markers show the simulation results with error bars representing ± one standard deviation. The beige marker indicates the result corresponding to the experimental true noise level.

From these results, we make three observations. First, at the experimental sensor noise level, there is no measurable effect on the learning rate (Supplementary 4 Fig. 1a), and minimal effect on the steady state error (Supplementary 4 Fig. 1b). Second, noise can affect adaptation performance. Higher noise standard deviations generally increase error and number of iterations. However, this requires increasing the noise level by several orders of magnitude for a noticeable effect. Third, despite large, imposed noise levels in the simulations, convergence is always achieved, underscoring the robustness of the model-free algorithm.

The discrepancies observed between FE predictions and the physical prototype (main text Fig. 3c–f) are therefore not due to measurement noise. Mechanical imperfections such as hinge friction, hinge imperfections, or slight out-of-plane motions are the probable contributors to the observed discrepancies. However, all of these are directly accounted for in the mechanics of the structure and hence automatically captured in physical learning.